# Title: The TikToking troll and weaponization of conscience: A systems perspective

**Authors:** Michelle Espinoza[1]*

**Affiliations:**

[1]Marymount University; Arlington, 22207, United States.

* Email: m0e73021@marymount.edu

**Abstract:** Cybercrime is a pervasive threat that impacts every facet of society. Its reach transcends geographic borders and extends far beyond the digital realm, often serving as the catalyst for offline crimes. As modern conflicts become increasingly intertwined with cyber warfare, the need for interdisciplinary cooperation to grasp and combat this escalating threat is paramount. This case study centers around a controversial TikToker, highlighting how weaponization of conscience can be leveraged to manipulate multiple actors within a propagandist's target population. Weaponization of conscience is a tactic used by fraudsters to camouflage their activity, deceive their victims, and to extend the effectiveness of their modi operandi. Research shows that 95% of cybersecurity incidents are the result of human error and 90%+ begin with a phishing attempt. Honing the capacity to identify and dissect strategies employed by fraudsters along with how individual reactions unfold in the larger system is an essential skill for organizations and individuals to safeguard themselves. Understanding cybercrime and its many interconnected systems requires examination through the lens of complexity science.

**Main Text:** Conscience is defined as an intrinsic motivational force guided by a person's moral beliefs. Those moral beliefs may or may not be rooted in religion and a person's moral beliefs can change or evolve. To weaponize something is to make it possible to use something to attack, manipulate, control, or influence a person or group. Weaponization of conscience in the realm of cybersecurity is a tool/tactic employed by fraudsters to camouflage their activity and extend the effectiveness of their modi operandi (Espinoza, 2024). When conscience is modeled as a complex system, tweaking subcomponents of the decision system can predictably alter the outcome. Propagandists use weaponization of conscience to influence their target's behaviors through information and emotion. Figure 1 shows a simplified illustration of the conscience system based on Green's (2009) dual process moral decision theory which posits that human moral judgment is influenced by two distinct processes: intuitive-emotional and deliberative. Dual process theory suggests that moral decisions are not the result of a single cognitive process but rather the interplay between two different types of processes (*1*). 95% of all cyber incidents are human-enabled, this is why more emphasis on identifying and understanding the complex interplay of group and individual decision-making theories is essential (*2*).

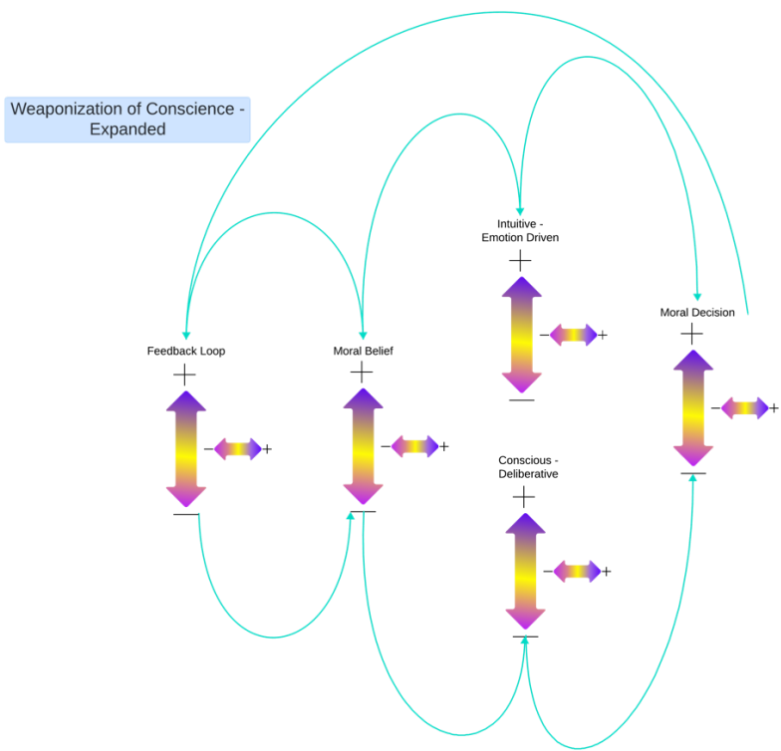

*Figure 1 Weaponization of Conscience Model*

Because these principles can be applied to multiple similar situations and to avoid giving the TikToker subject more notoriety, they will be referred to as JT. This researcher does not seek to modify readers' individual belief systems, but rather to provide a lens for analyzing how conscience may be weaponized to manipulate the dynamics between actors in a complex system, and how the reactions may fit into the larger social system in which they are embedded.

State-sponsored internet trolls primarily seek to manipulate public opinion, disseminate propaganda, or sow discord among their target population to further the interests of their sponsoring government. Internet trolls are not a monolithic group, and their tradecraft is ever evolving. A notable aspect of contemporary state-sponsored online manipulation efforts is the use of companies posing as authentic local news sources to propagate misinformation (*3, 4*). The Toronto-based watchdog group known as Citizen Lab recently discovered more than 100 websites in 30 countries posing as local news outlets. Citizen Lab's investigation was prompted by South Korea's National Cyber Security Center's report in November 2023 which revealed 18 news sites linking back to a Chinese public relations firm (*5*).

Fraudulent news outlets are particularly effective for exploiting data voids (*6*). When we read a sensational news headline, our inclination is to search online for updates or more in-depth information on the story. In cases where the story is untrue, then there will not be competing narratives from credible news outlets returned in the search results. This data void paves the way for a deluge of fabricated social media profiles and news outlets to disseminate the fictitious news story without any hindrance or opposition. A less-overt tactic might be for the propagandist to exploit confirmation bias by embellishing a news story with untrue or irrelevant details that subtly reinforce existing beliefs.

The Internet Research Agency (IRA) is another example of a state-sponsored industrialized troll farm dedicated to online propaganda and misinformation campaigns. The group, based out of St. Petersburg, were known to have specialized in online influence operations by creating legions of fictitious social media profiles that promoted opposing viewpoints and exacerbated societal divisions in the United States. The IRA even posed as real activist groups and staged protests in multiple countries to further fuel the flames of polarization (*7*). Using stolen identities, the IRA opened multiple payment accounts under American names and purchased ads to further their propaganda efforts (*8*). Figures 2 and 3 illustrate how the IRA combined weaponization of conscience with group polarization theory to manipulate the desired outcome.

Group polarization theory suggests that when a homogenous group hold similar opinions, group members with opinions or information that contradicts the emerging consensus are less likely to share the dissenting information. Consequently, the group becomes firmer entrenched in their belief and move towards a more extreme position from where they were prior to the debate (*9*). This firmer entrenchment increases resistance to new ideas when presented with evidence that challenges existing beliefs, a phenomenon known as the Semmelweis reflex. Building on McGuire's (1964) inoculation theory, Bizer and Petty (2005) found that by framing a preference as being in opposition to something, instead of in support of something, a person's preference became more resistant to persuasion (*10*).

In-depth knowledge of decision-making theories, though helpful, is not a prerequisite for individuals to begin identifying and analyzing how fraudsters' tactics may elicit certain reactions and how those reactions unfold within a larger system. The human mind is generally poor at interpreting multiple-loop nonlinear feedback systems without visualizing the system's interactions (*11*). Specialized applications exist for modeling dynamic systems, but the average person can gain considerable insight into a system's dynamics by mapping the conceptual system on a piece of paper.

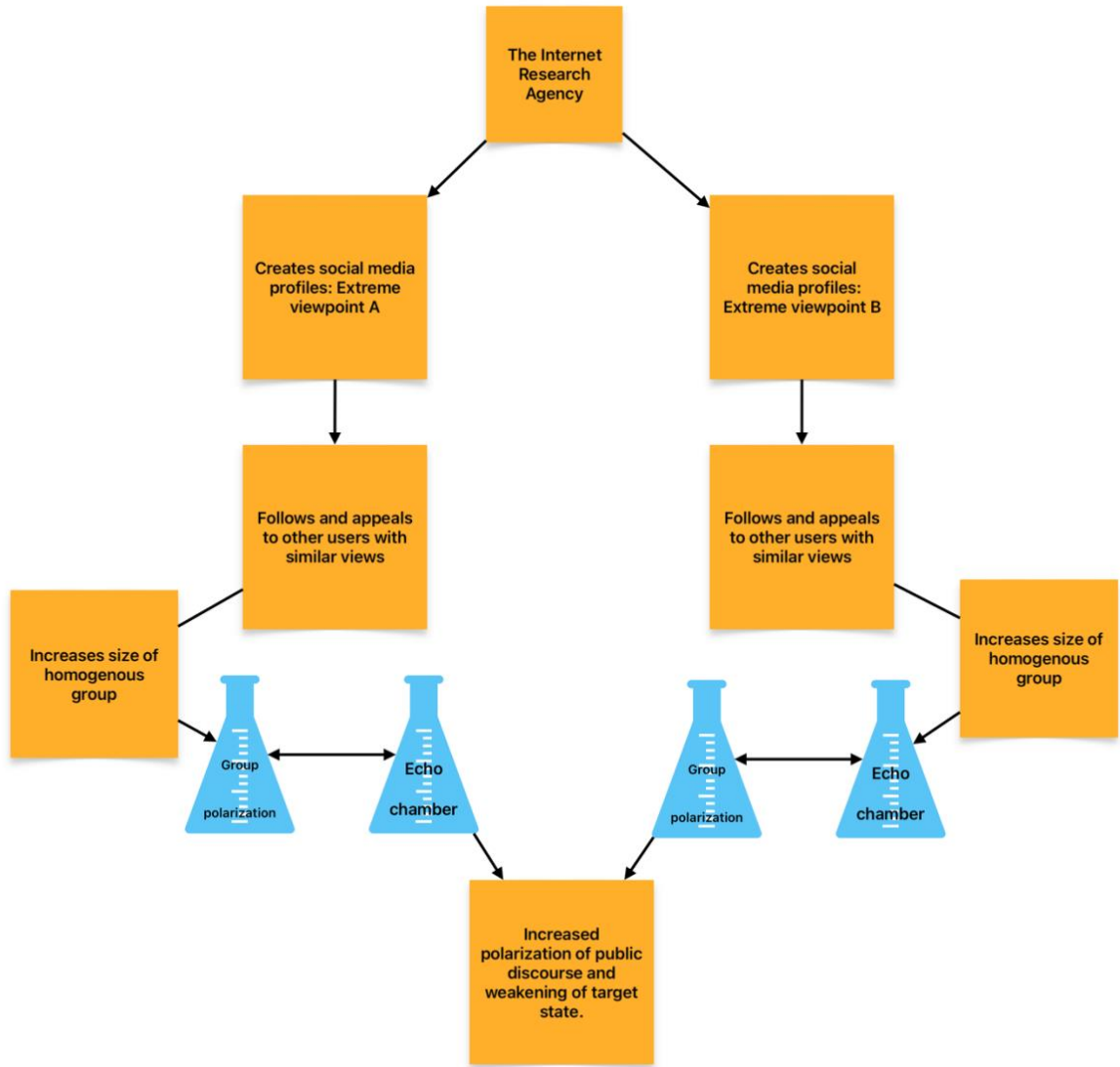

*Figure 2 Example of how the Internet Research Agency manipulated public discourse*

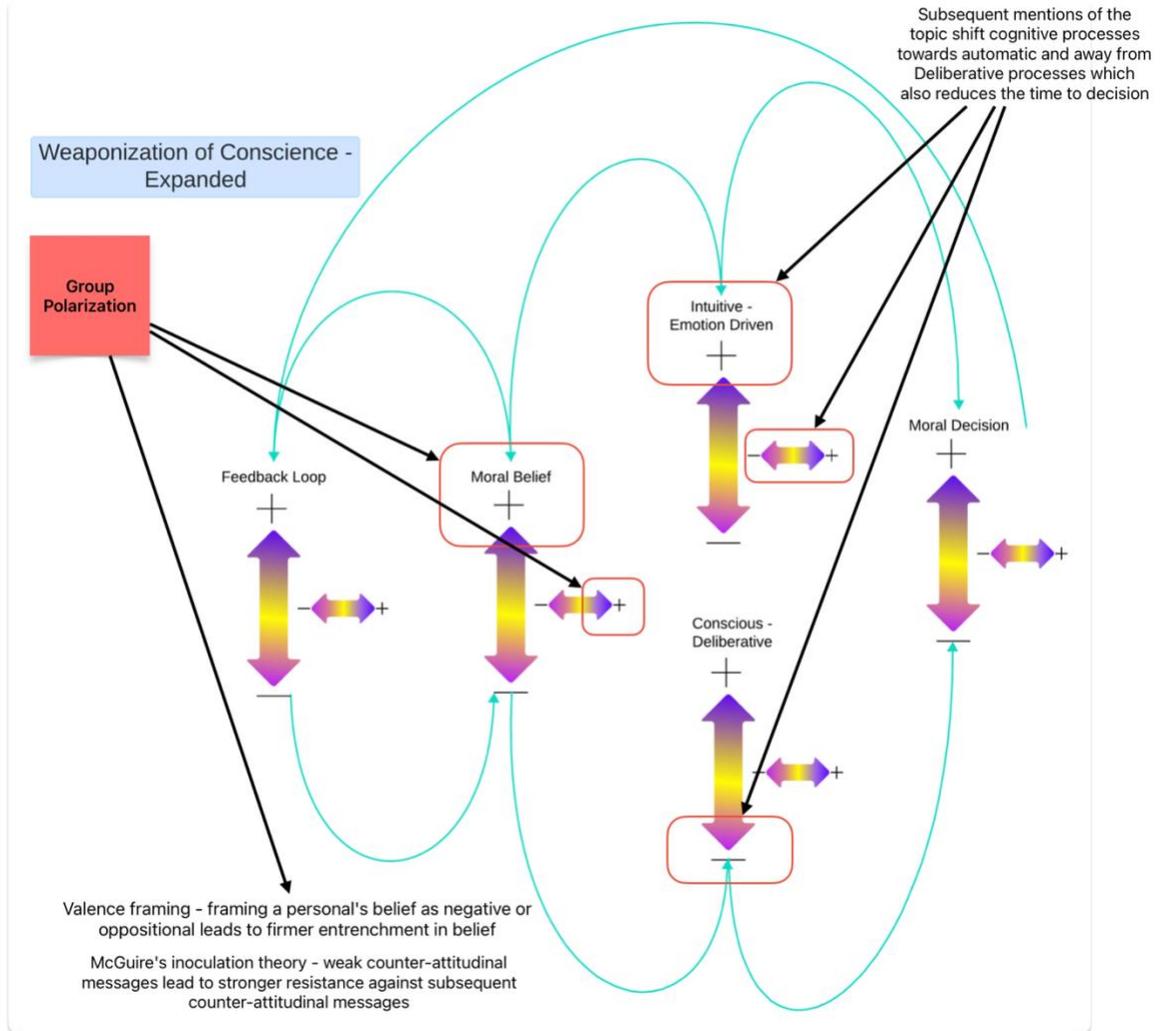

*Figure 3 Weaponization of Conscience using group polarization*

**JT The TikToker Troll**

JT is known for controversial TikTok videos where he urges his compatriots in Venezuela to embark on a journey to the United States, boasting that freeloading is easy and that he enjoys a luxurious lifestyle with relatively little work. In JT's videos, he delivers incendiary statements suggesting to migrants that he makes thousands of dollars panhandling, exploits government assistance programs, and more recently advertised that migrants can obtain a free house by invoking squatter's rights. The diagram in Figure 4 depicts an expanded model of the system of reactions that JT's videos might elicit. This article focuses on a small selection of theories and concepts stemming from the model.

Complex system models incorporate assumptions and estimations about how subsystems interact. As more or better information becomes available, models can be revised or expanded. When constructing these models with the aim of finding a solution, it is important to note that in complex systems our attention most often gravitates to the point at which intervention will fail (*11*). Thinking of the model in terms of a treasure map, rather than x marking the location of buried treasure, consider the marked spot to be the starting point for examining the surrounding system interactions. Another useful analogy are canaries in the coal mines. Canaries were once used as early warning systems for miners to detect dangerous levels of carbon monoxide gas. Reviving the canary would not reduce carbon monoxide levels. Rather, it would prolong the miners' exposure to rising carbon monoxide levels potentially killing the miners and the canary in the process.

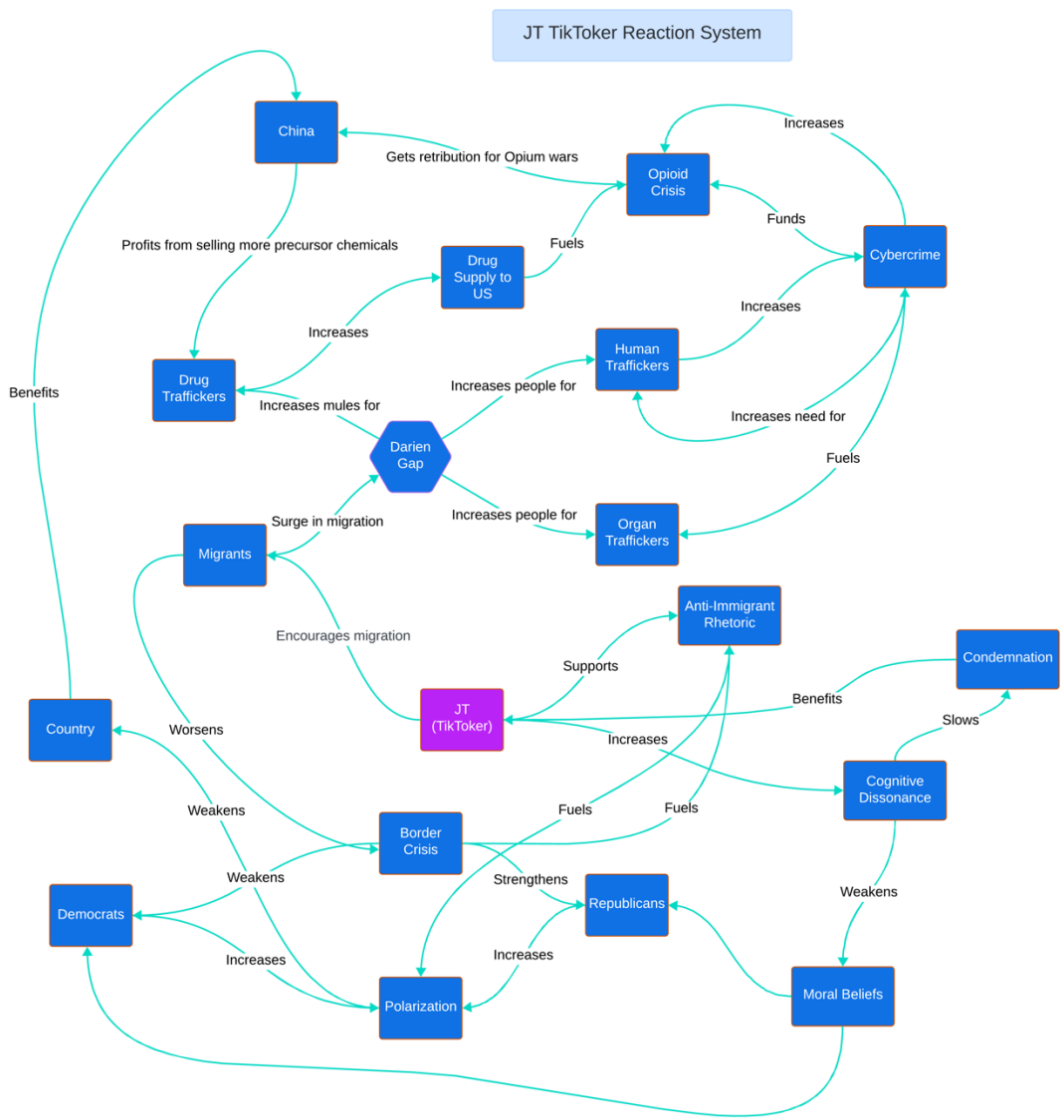

*Figure 4 Hypothetical model of system reactions to JT's videos*

**Information Cascades and the Darién Gap**

Information cascades refer to situations where individuals make decision based on the observed actions of others, sometimes without critical analysis or despite personal knowledge that counters the logic and soundness of the decision being made. We see this phenomenon in financial markets, discount shopping, and product selection where shoppers may pick one product over another because the product has more reviews.

JT's videos, particularly the suggestion that every migrant can simply take ownership of vacant homes, encourages a surge in migration. As noted in a US congressional panel last year, misinformation is known to contribute to migration surges (*12*). Media coverage of migration surges exacerbates the information cascade effect, leading more migrants to risk the dangerous trek North. Misinformation is disseminated at a faster rate than truth, particularly when the misinformation is more novel, unbelievable, or sparks intense emotional reactions (*13*, *14*). This creates a feedback loop as depicted in Figure 5 where more people learn about the opportunity to obtain a free home, which could lead to a surge in migration. The surge in migration leads to more media coverage which reduces deliberation processes.

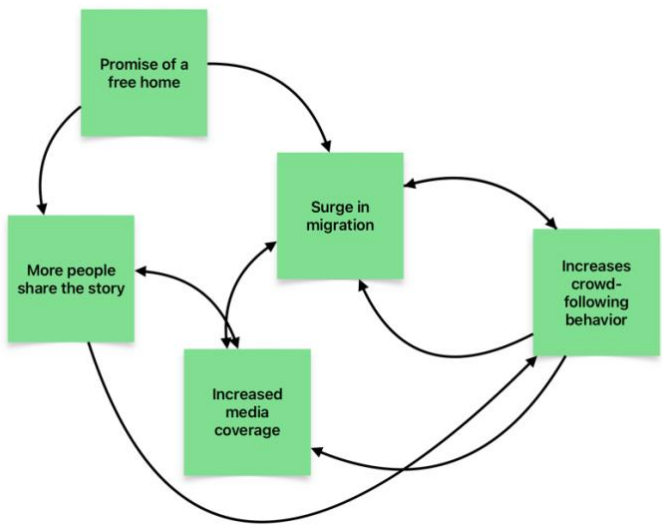

*Figure 5 Feedback loop migration surge*

To provide a counterexample, a headline such as 'vegetables are good for you' is unlikely to spark a rush on grocery store produce aisles.

The Darién gap is the 60-mile stretch of jungle and treacherous terrain connecting Colombia and Panama. Often referred to as the deadliest migrant path in the world, it is controlled by armed guerillas, drug traffickers, and human trafficking networks. The lawlessness of the region coupled with lack of infrastructure make it nearly impossible to provide aid to the thousands of migrants who perish along the path each year. Migrants who are unable to pay for safe passage are sometimes given the option of muling drugs for traffickers. Considered disposable, and [essentially] unwanted by any country, the migrants transiting the Darién are vulnerable to exploitation by human and organ traffickers. A surge in migrants to the deadly Darién can spark the same predatory instinct that drives bears to the annual salmon run in Alaska. Conversely, a large drug seizure by law enforcement could spark the need for additional drug mules and serve as an incentive for traffickers to attempt triggering a migration surge through disinformation.

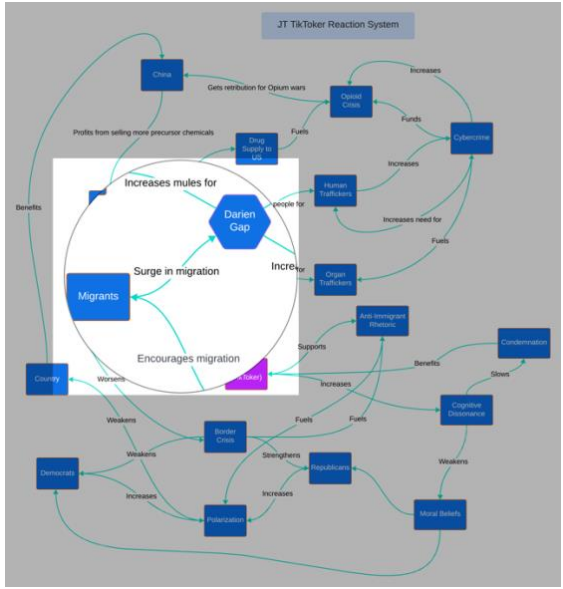

*Figure 6 Magnified view of the Darien Gap and migration surge referenced in the system*

## Cognitive Dissonance

Cognitive dissonance is the mental discomfort that arises when a person holds two or more conflicting beliefs, values, or attitudes. When there is inconsistency between a person's behavior and beliefs or between their beliefs and new information, people employ a variety of strategies to reduce the mental discomfort such as modifying their moral beliefs and increasing rationalization. Cognitive dissonance may also shift decision-making more towards deliberative processes and extend the time between deliberation and decision-making.

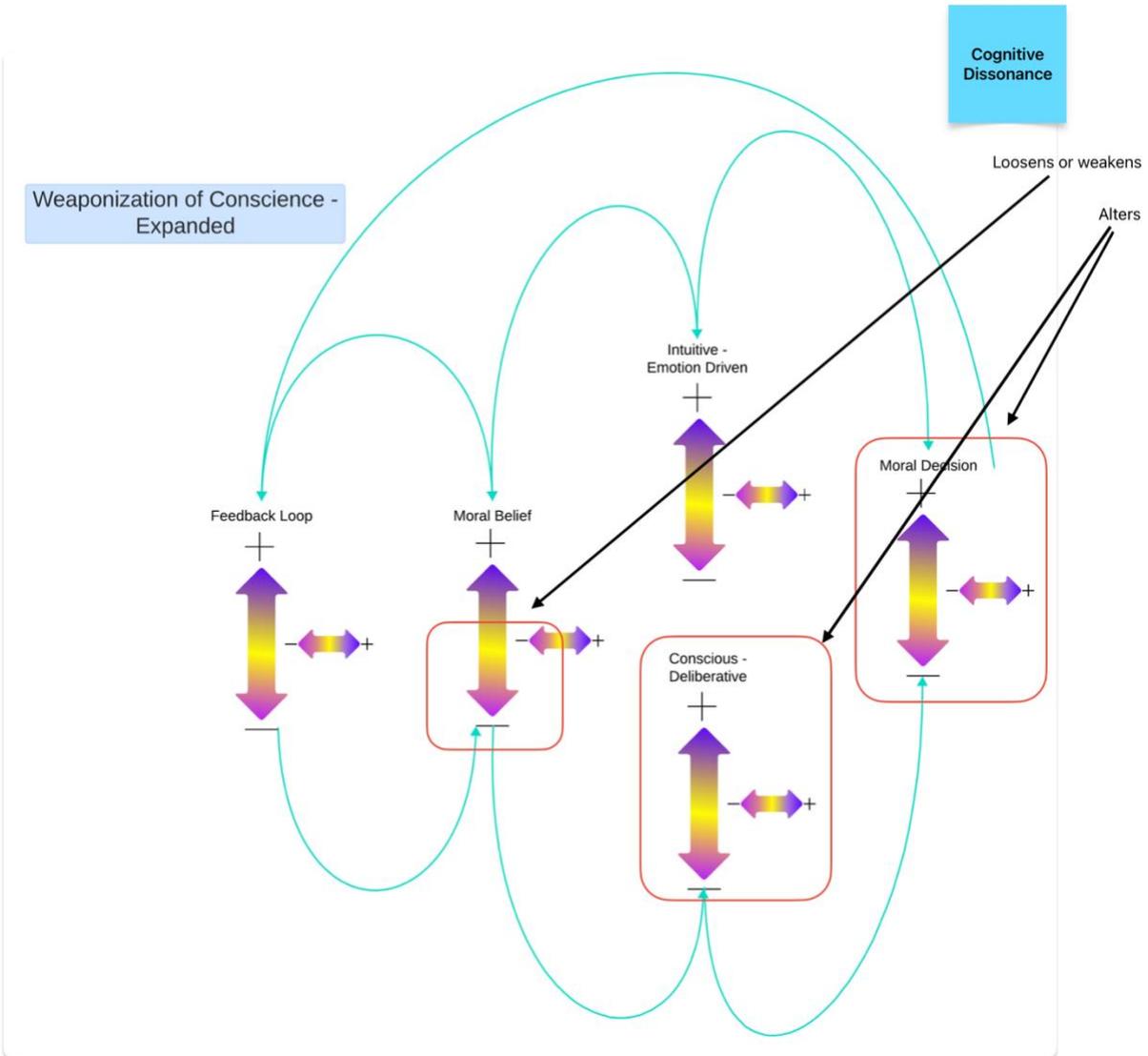

*Figure 7 Cognitive dissonance in weaponization of conscience*

Though not empirically tested, the cognitive dissonance created by JT's videos and statements appeared to slow condemnation, which extended the effectiveness of his tactics. JT posts multiple videos holding an infant who he claims is his daughter and states that the sympathy his daughter elicits from strangers increases the money JT receives when panhandling. JT also thanks President Biden in his videos and tells compatriots that all the rumors of the American Dream are true, insinuating that the American Dream is generally described as exploitation of a welfare state.

Consider how the following conflicting statements may deepen existing societal divisions or create fractures where solidarity exists, or even create solidarity where divisions exist. How might individuals react to reduce cognitive dissonance?

Conflicting Set A:

1. I feel a sense of solidarity and brotherhood with my compatriots.

2. I condemn JT, he is an embarrassment to our country.

Conflicting Set B:

1. I feel compassion for migrant families fleeing unbearable conditions in search of a better life here in the U.S.

2. I do not feel compassion for JT, he and his family are freeloaders taking advantage of the system.

Conflicting Set C:

1. I have not condemned the occupation or colonization of land by others.

2. Land-squatting is wrong, and I condemn JT's suggestion to invoke squatter's rights.

Conflicting Set D:

1. We are a nation of laws; we must support the rule of law.

2. JT is using our laws against us; we should not support these laws.

**Conclusion**

JT's TikTok infamy was short-lived, and his antics were largely dismissed or condemned by his compatriots online, but state-sponsored actors and other fraudsters continue operating in our midst, undetected and costing governments, organizations, and individuals trillions of dollars per year.

Perhaps the most severe consequence of continued weaponization of conscience will not be financial, but rather that we grow cold, cynical, and indifferent to each other's humanity. The collective impact of each individual's choices shape the dynamics of the system which connects us all.

**Acknowledgments:** NA

   **Funding:** NA

   **Competing interests:** Authors declare that they have no competing interests.

   **Data and materials availability:** All data are available in the main text or the supplementary materials.

**Supplementary Materials:**

NA